\documentclass[12pt,preprint]{aastex}

\def\h2{H$_2$~}

\makeatother

\begin{document}

\slugcomment{Accepted by ApJL}
\title{Molecular Hydrogen in a Damped Lyman-$\alpha$ System
at $z_{\rm abs}=4.224$}

\author{C. Ledoux\altaffilmark{1}, P. Petitjean\altaffilmark{2,3},
R. Srianand\altaffilmark{4}
}

\begin{abstract}

We present the direct detection of molecular hydrogen at the highest redshift
known today ($z_{\rm abs}=4.224$) in a Damped Lyman-$\alpha$ (DLA) system
toward the quasar PSS J\,1443$+$2724. This absorber is remarkable for having
one of the highest metallicities amongst DLA systems at $z_{\rm abs}>3$, with a
measured iron abundance relative to Solar of $-1.12\pm 0.10$. We provide for
the first time in this system accurate measurements
of N\,{\sc i}, Mg\,{\sc ii}, S\,{\sc ii} and Ar\,{\sc i} column densities.
The sulfur and nitrogen abundances relative to Solar, $-0.63\pm 0.10$
and $-1.38\pm 0.10$ respectively, correspond exactly to the primary nitrogen
production plateau. H$_2$ absorption lines are detected in four
different rotational levels (${\rm J}=0$, 1, 2 and 3) of the vibrational
ground-state in three velocity components with total column densities
of $\log N({\rm H}_2)=17.67$, 17.97, 17.48 and 17.26 respectively. The
${\rm J}=4$ level is tentatively detected in the strongest component
with $\log N({\rm H}_2)\sim 14$. The mean molecular fraction
is $\log f=-2.38\pm 0.13$, with
$f=2N({\rm H}_2)/(2N({\rm H}_2)+N($H\,{\sc i}$))$. We also
measure $\log N($HD$)/N($H$_2)<-4.2$. The excitation temperatures $T_{01}$ for
the two main components of the system are 96 and 136~K respectively. We
argue that the absorbing galaxy, whose star-formation activity must
have started at least $2-5\times 10^8$~yrs before $z=4.224$, is in a
quiescent state at the time of observation. The density of the gas is small,
$n_{\rm H}\le 50$~cm$^{-3}$, and the temperature is of the order
of $T\sim 90-180$~K. The high excitation of neutral carbon in one of
the components can be explained if the temperature of the Cosmic Microwave
Background Radiation has the value expected at the absorber redshift,
$T=14.2$~K. These observations demonstrate the feasibility to study H$_2$
at the highest redshifts provided high enough spectral resolution and
good S/N ratio are achieved.

\end{abstract}



\altaffiltext{1}{European Southern Observatory,
Alonso de C\'ordova 3107, Casilla 19001, Vitacura, Santiago, Chile;
E-mail: cledoux@eso.org}
\altaffiltext{2}{Institut d'Astrophysique de Paris -- CNRS,
98bis Boulevard Arago, F-75014 Paris, France;
E-mail: petitjean@iap.fr}
\altaffiltext{3}{LERMA, Observatoire de Paris,
61 Avenue de l'Observatoire, F-75014, Paris, France}
\altaffiltext{4}{IUCAA, Post Bag 4, Ganesh Khind, Pune 411\,007, India;
E-mail: anand@iucaa.ernet.in}

\section{Introduction}

Although molecular hydrogen, H$_2$, is the most abundant mo\-le\-cu\-le in the
Universe, it is very difficult to detect directly. Emission lines from the
tracer molecule CO are usually used instead but, at $z>4$, only the
brightest objects can be detected with current facilities
(e.g., Omont et al. 1996; Solomon \& Vanden Bout 2005). For the time being,
the only way to detect H$_2$ directly at high redshift is to search for the
absorption signature of the molecular UV transition lines from
Damped Lyman-$\alpha$ (DLA) systems observed in the spectra of quasars or
those of $\gamma$-ray burst afterglows.

DLA systems are characterized by very large neutral hydrogen column
densities, $N($H\,{\sc i}$)\ga 10^{20}$ cm$^{-2}$, that are typical of
sightlines through local spiral galaxies. Inferred metallicities typically
vary between [Zn,S/H$]=-2$ and -0.5 at $2\la z_{\rm abs}\la 3$ and are smaller
at larger redshifts (e.g., Prochaska et al. 2003). The dust content is less
than or about 10\% of what is observed in the Galactic ISM for
similar H\,{\sc i} column densities (Ledoux et al. 2003). The detection of
H$_2$ molecules in high-redshift DLA systems through Lyman-Werner band
absorption is observationally challenging due to the presence of the Ly$\alpha$
forest. High spectral resolution, good S/N ratio and large wavelength
coverage are simultaneously needed to detect the H$_2$ lines and avoid
spurious detections (e.g.,
Levshakov \& Varshalovich 1985; Ge \& Bechtold 1999; Petitjean et al. 2000; Levshakov et al. 2002; Ledoux et al. 2002, 2003; Cui et al. 2005).

We are conducting a survey for molecular hydrogen in DLA systems at
high redshift ($z_{\rm abs}>1.8$; see Ledoux et al. 2003), using the
Ultraviolet and Visible Echelle Spectrograph (UVES; Dekker et al. 2000) at
the European Southern Observatory's Very Large Telescope, down to a
detection limit of typically $N({\rm H}_2)=2\times 10^{14}$ cm$^{-2}$. Out
of the 75 systems observed up to now, 14 have firm detections of associated
H$_2$ absorption lines. In this Letter, we present the highest redshift
detection in our sample, in the $z_{\rm abs}=4.224$ DLA system toward
the quasar PSS J\,1443$+$2724. We describe the observations in Sect.~2 and the
analysis of the data in Sect.~3. We conclude in Sect.~4.

\section{Observations}

The quasar PSS J\,1443$+$2724 ($z_{\rm em}=4.42$, $r\sim 19.3$) was observed
with UVES in service mode on March 2-3 and 17-19, 2004. The red
spectroscopic arm of UVES was used in the standard configuration with
the central wavelength adjusted to 580~nm. The wavelength coverage
was 478$-$681~nm with only a small gap, between 575 and 584~nm,
corresponding to the physical gap between the two red arm CCDs. The CCD pixels
were binned $2\times 2$ and the slit width was fixed to $1\arcsec$,
yielding, under the typical $0\farcs 7$ seeing conditions achieved during the
observations, a resolving power $R\approx 48000$. The total on-source
integration time was 7.3~h, split into five exposures. The data were
reduced using the UVES pipeline (Ballester et al. 2000) which is available as
a dedicated context of the ESO MIDAS data reduction system.
The wavelength scale of the spectra
reduced by the pipeline was converted to vacuum-heliocentric values
and individual exposures were scaled, weighted and combined altogether to
produce the final object flux and variance spectra.

\section{Analysis}

\subsection{The high metallicity and the star-formation history}

The metallicity of the DLA system toward PSS J\,1443$+$2724 is the
highest among the ten $z>4$ DLA systems investigated up to now
(Prochaska et al. 2003). Total integrated column densities and metallicities
for N\,{\sc i}, Mg\,{\sc ii}, S\,{\sc ii}, Ar\,{\sc i} and Fe\,{\sc ii},
obtained from Voigt-profile fitting of the observed transition lines, are given
in Table~\ref{Tabmetal}. With the exception of Fe\,{\sc ii}, these ions are
observed here for the first time. We derive metallicities relative to Solar
of [S/H$]=-0.63\pm 0.10$ and [Fe/H$]=-1.12\pm 0.10$. The latter value is
consistent with the previous measurement by Prochaska et al. (2001) but we
note that the newly determined H\,{\sc i} and Fe\,{\sc ii} column densities
are both larger by about 0.15~dex. Even though the depletion factor is moderate
(i.e., [S/Fe$]=+0.49$), because of the high metallicity the dust-to-gas
ratio measured in this system ($\log\kappa =-0.8$) is similar to that of other
systems in which H$_2$ is detected (see Fig.~15
of Ledoux et al. 2003). Substantial star-formation activity must have taken
place in the past history of this DLA system. Note that the age of the Universe
at $z=4.22$ is 1.5~Gyr (for $\Omega_\Lambda =0.73$, $\Omega_{\rm m}=0.27$ and
H$_0=71$ km s$^{-1}$ Mpc$^{-1}$).

Nitrogen is thought to have both primary and secondary origins depending on
whether the seed carbon and oxygen nuclei are produced by the star itself or
are already present in the ISM from which the star formed. In the case
of secondary nitrogen production, the ratio of nitrogen to oxygen
abundances increases with increasing oxygen abundance, whereas for
primary production this ratio remains constant and nitrogen tracks oxygen
(this is the so-called primary plateau). In
the low-metallicity H\,{\sc ii} regions of nearby galaxies, [N/O] abundance
ratios tend to lie around the primary plateau (with [N/O$]=-0.73\pm 0.13$) for
[O/H$]<-1$, and have a secondary behaviour for [O/H$]>-1$
(van Zee et al. 1998; Izotov \& Thuan 1999). Similarly, most DLA systems show a
plateau at [N/O$]\approx -0.9$ (see
Centuri\'on et al. 2003; Pettini et al. 2002). The abundance of nitrogen
relative to sulfur at $z_{\rm abs}=4.224$ toward
PSS J\,1443$+$2724, [N/S$]=-0.75$, corresponds exactly to the primary nitrogen
production plateau (assuming [O/S$]=0$). If we assume the widely accepted
hypothesis that primary nitrogen is produced in intermediate-mass stars
($4\leq M_\odot\leq 8$) during the Asymptotic Giant Branch phase
(Henry et al. 2000), then star-formation activity must have started in
this system at least $2-5\times 10^{8}$~yrs before $z=4.224$ accounting for the
characteristic lag time for intermediate-mass stars to eject nitrogen. This is
similar to what is observed in the $z_{\rm abs}=4.383$ system toward
BRI\,1202$-$0725 (D'Odorico \& Molaro 2004).

Argon can be ionised in H\,{\sc i} regions and is therefore a tracer of
the radiation field. In the presently studied system, we find
[Ar/S$]=-0.30$, which is higher than in most other DLA systems
(Vladilo et al. 2003). This is consistent with the associated galaxy being in a
quiescent state at the time of observation, as also inferred below from
the analysis of H$_2$ lines.

\subsection{Molecular hydrogen}

\begin{figure}
\centering
\vbox{
\centering
\includegraphics[width=8.5cm,bb=60 320 403 769,clip,angle=0]{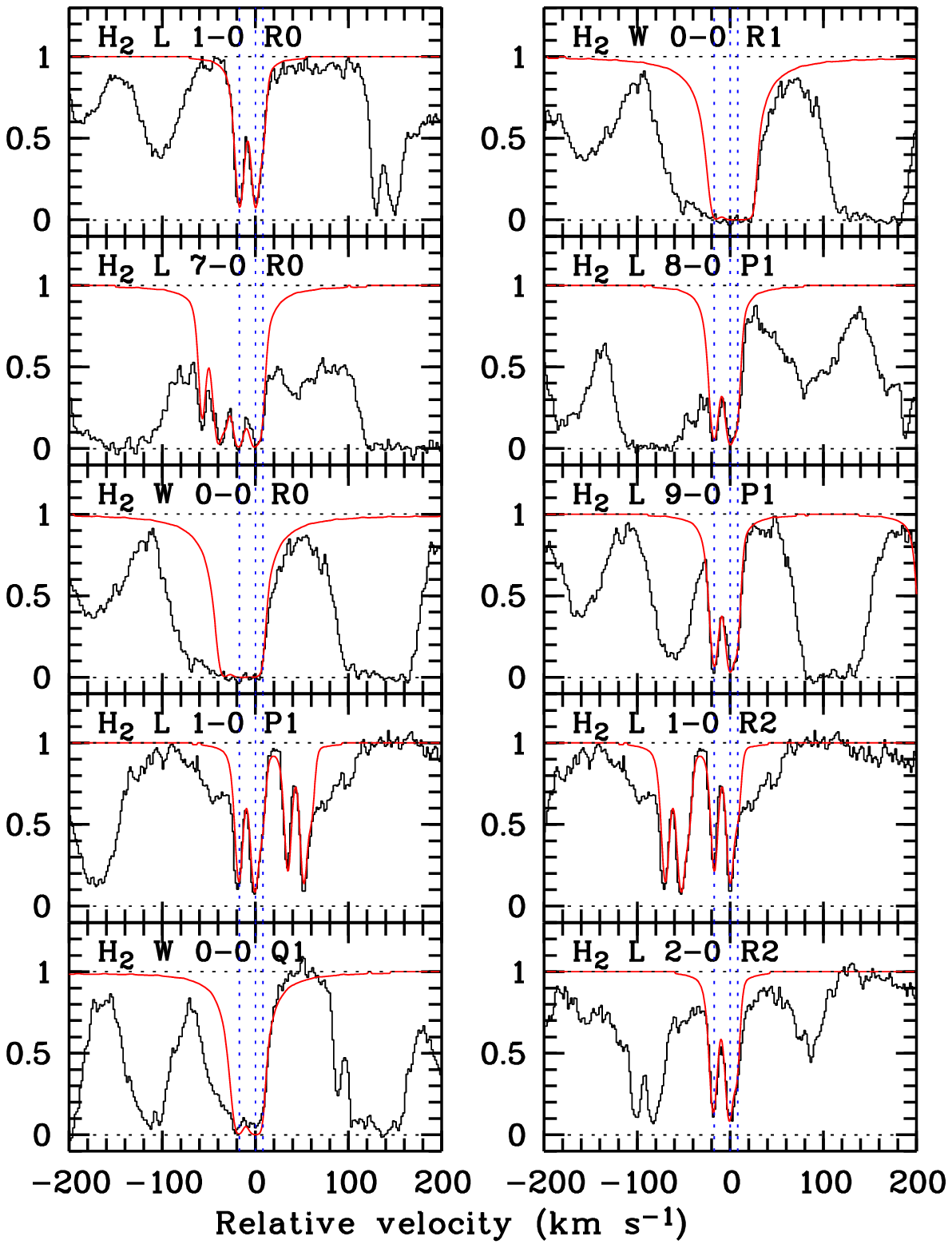}
\includegraphics[width=8.5cm,bb=60 320 403 769,clip,angle=0]{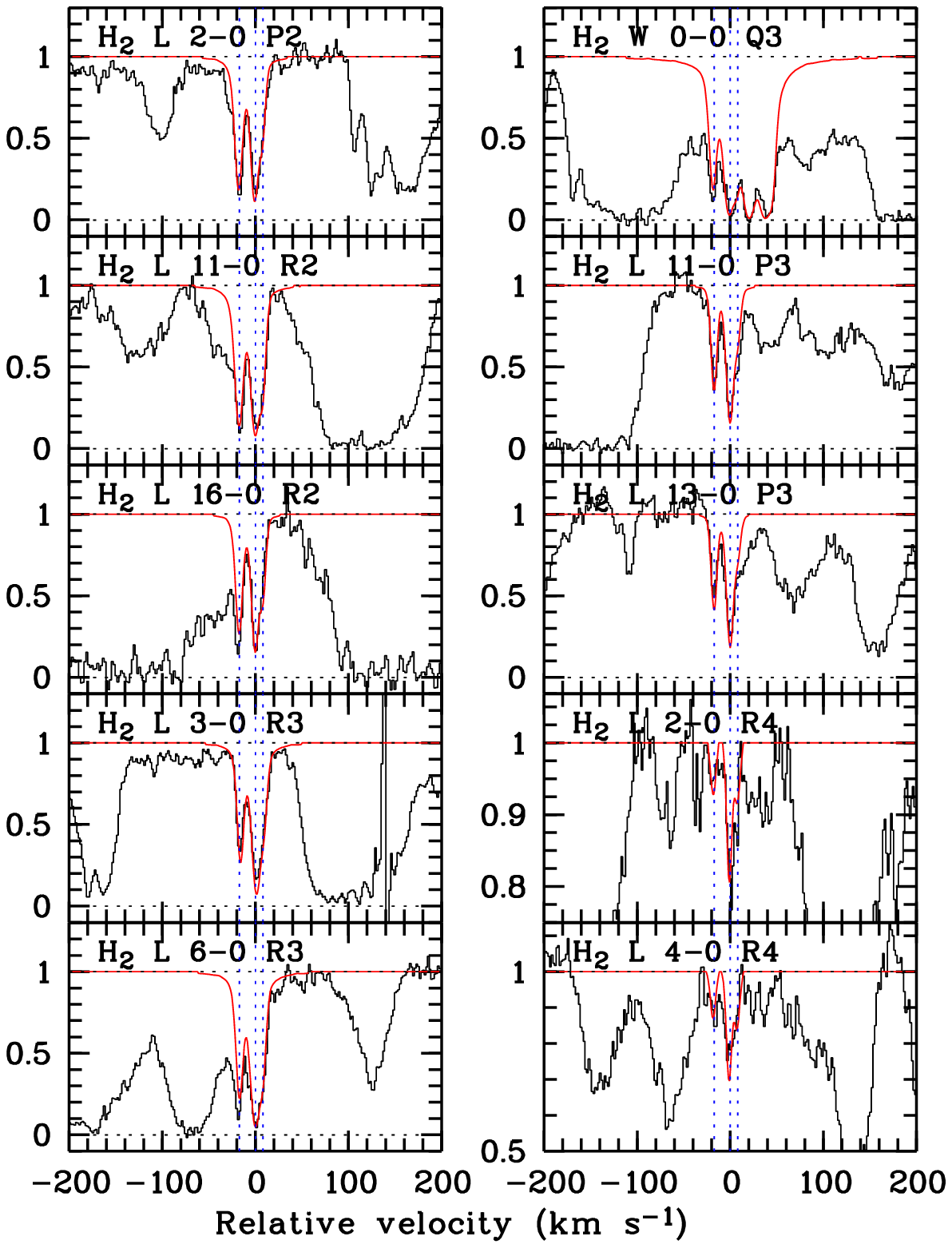}}
\caption[]{Velocity profiles of selected transition lines from the ${\rm J}=0$,
1, 2, 3 and 4 rotational levels of the vibrational ground-state Lyman and
Werner bands of H$_2$ at $z_{\rm abs}=4.224$ toward PSS J\,1443$+$2724. The
best Voigt-profile fit is superposed on the observed spectrum with vertical
dotted lines marking the location of the three detected components.
}
\label{Fig1}
\end{figure}

The H$_2$ transitions are numerous and the spectral resolution of our data is
high enough to allow unambiguous detection and detailed analysis of H$_2$
(see Fig.~\ref{Fig1}). Great care has been exercised when fitting
the absorption lines with multi-component Voigt profiles. H$_2$ is detected
in four different rotational levels (${\rm J}=0$, 1, 2 and 3) of
the vibrational ground-state in three velocity components with total
column densities of $\log N({\rm H}_2)=17.67$, 17.97, 17.48 and 17.26
respectively (see Tables~\ref{Tabmetal} and \ref{TabH2}). The ${\rm J}=4$ level
is tentatively detected in the strongest component with
$\log N({\rm H}_2)\sim 14$. Blending is heavy however and this could be an
upper limit. The detection of H$_2$ confirms the previous claim that most of
the gas in this DLA system is cold (Howk et al. 2005). The total H$_2$
column density is one of the largest observed in DLA systems and the molecular
fraction is $\log f=-2.38\pm 0.13$,
with $f=2N({\rm H}_2)/(2N({\rm H}_2)+N($H\,{\sc i}$))$. HD is not detected
down to $3\sigma$ limits of $\log N($HD$)<13.6$ and 13.7 for ${\rm J}=0$ and 1
respectively, implying that $[$HD$/$H$_2]<-4.2$.

The excitation temperatures $T_{01}$ for the two main components of the system
are 96 and 136~K respectively (see Table~\ref{TabH2}). The
excitation temperatures for higher rotational levels are slightly
larger, possibly as a consequence of the levels being populated by radiative
or formation pumping. By equating H$_2$ formation and destruction rates, we
write:
\begin{equation}
SIn({\rm H}_2) = Rn({\rm H})n
\end{equation}
where $n=n({\rm H})+2n({\rm H}_2)$ is the particle density, $I$ is the H$_2$
dissociation rate in the unshielded UV background (in s$^{-1}$) and
$S=[N({\rm H}_2)/10^{14}\ {\rm cm}^{-2}]^{-0.75}$ is the correction factor for
shielding (Draine \& Bertoldi 1996; Hirashita \& Ferrara 2005).
$R=4.1\times 10^{-17}[T/10^2\ {\rm K}]^{0.5}(Z/Z_\odot )D$ cm$^3$ s$^{-1}$ is
the H$_2$ formation rate per unit volume and time, $Z/Z_\odot =0.23$ is the
metallicity relative to Solar and $D=0.7$ is the fraction of metals into dust.

The H$_2$ ${\rm J}=4$ level is mostly populated by pumping of lower rotational
levels by absorption of UV photons from the ambient radiation field
and subsequent radiative cascade, and by direct formation of the molecule in
this excited state. Following the simple procedure introduced
by Jura (1975), we write the equation describing the equilibrium of the
${\rm J}=4$ level population as:
\begin{equation}
9.1p_{4,0}SIn({\rm H}_2,{\rm J}=0)+0.19Rn({\rm H})n~=~A(4\rightarrow 2)n({\rm H}_2,{\rm J}=4)
\end{equation}
where $p_{4,0}=0.26$ is the pumping efficiency into the ${\rm J}=4$ level from
the ${\rm J}=0$ level
and $A(4\rightarrow 2)=2.8\times 10^{-9}$ s$^{-1}$ denotes the spontaneous
transition probability from ${\rm J}=4$ to ${\rm J}=2$.

Combining the above equations and using the total column densities given in
Table~\ref{Tabmetal}, we find $Rn=4\times 10^{-16}$ s$^{-1}$. This is
probably an upper limit given the possibility that $N({\rm J}=4)$ be an upper
limit. Assuming a temperature of $T=150$~K, we derive a density
of $n_{\rm H}\leq 50$ cm$^{-3}$. This value could be slightly larger in case
of any $\alpha$-element enhancement. The corresponding H$_2$ dissociation
rate is $I\leq 3\times 10^{-10}$ s$^{-1}$. This is of the order of or
smaller than the highly variable photo-dissociation rate observed in the ISM
of the Galaxy (e.g., Hirashita \& Ferrara 2005). This probably indicates,
as expected above, that there is little on-going star formation in the galaxy
associated with this DLA system. This is consistent with the fact that
no emitting counterpart has been found using the Lyman-break techniques down
to $L\sim L^\star_{\rm LBG}/4$ (Prochaska et al. 2002) or in Lyman-$\alpha$
emission down to $2\times 10^{-17}$ erg cm$^{-2}$ s$^{-1}$ from 2D-spectroscopy
at the Canada-France-Hawaii Telescope (Ledoux 1999).

\subsection{The Cosmic Microwave Background temperature}

\begin{figure}
\centering
\includegraphics[width=8.5cm,bb=59 566 394 769,clip,angle=0]{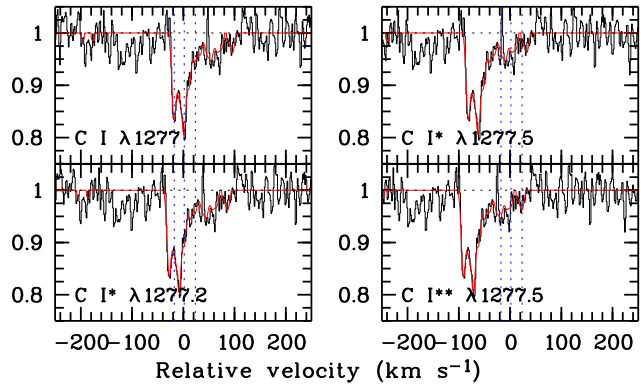}
\caption[]{Velocity profiles of C\,{\sc i} and C\,{\sc i}$^\star$ transition
lines at $z_{\rm abs}=4.22379$ and 4.22413 toward PSS J\,1443$+$2724. The
best Voigt-profile fit is superposed on the observed spectrum with vertical
dotted lines marking the location of the two detected components of the
system and a third, possible component.}
\label{Fig2}
\end{figure}

\begin{figure}
\centering
\includegraphics[width=8.5cm,bb=18 162 577 695,clip,angle=0]{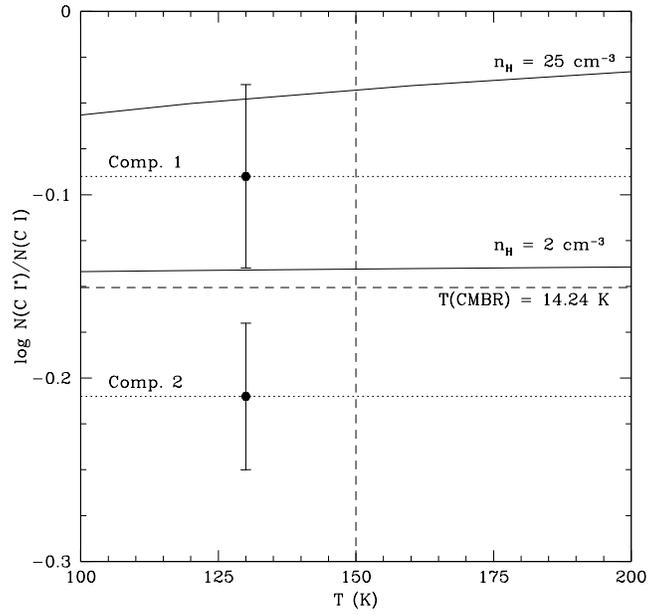}
\caption[]{$[$C\,{\sc i}$^\star /$C\,{\sc i}$]$ ratio versus temperature.
Filled circles with error bars are the measurements in the
$z_{\rm abs}=4.22379$ and 4.22413 components toward PSS J\,1443$+$2724.
LTE predictions for two different values of the particle density are shown
by solid lines. The value expected for excitation by the CMBR field alone
is shown by the horizontal long-dashed line.}
\label{Fig3}
\end{figure}

Absorption lines from the ground-state and first fine-structure excited level
of C\,{\sc i} are detected (see Fig.~\ref{Fig2}) at redshifts
of $z_{\rm abs}=4.22379$ and 4.22413, slightly larger than those of the
molecular components. The Doppler parameters are larger as well indicating
that the C\,{\sc i} gas spans a slightly larger velocity range than
the H$_2$ gas (see also Ledoux et al. 2003). We measure
$[$C\,{\sc i}$^\star /$C\,{\sc i}$]=-0.09\pm 0.05$ and $-0.21\pm 0.04$ in each
of the two components respectively. This is surprisingly small at
high redshift. Indeed, the value expected for excitation by the
Cosmic Microwave Background Radiation (CMBR) field alone, with a temperature of
14.2~K at $z=4.22$, is $-0.15$ (Silva \& Viegas 2002). In addition, part of the
excitation can be due to UV or IR pumping although this is probably negligible
here. In Fig.~\ref{Fig3}, we plot the expected values for
$[$C\,{\sc i}$^\star /$C\,{\sc i}$]$ as a function of temperature of the
cold gas. The first C\,{\sc i} component has a density
($n_{\rm H}\le 25$ cm$^{-3}$) that is similar to what is measured in the H$_2$
gas and is thus probably predominantly associated with it. The
second C\,{\sc i} component probably has a lower density and the corresponding
$[$C\,{\sc i}$^\star /$C\,{\sc i}$]$ ratio is consistent with what is expected
for excitation by the CMBR field alone.

\section{Conclusion}

The direct detection of H$_2$ at high redshift is a unique tool for
studying the ISM in the remote Universe. We have detected molecular hydrogen
in one of the highest redshift DLA systems known to date ($z_{\rm abs}=4.224$).
We have shown that the gas is probably the left-over of
intense star-formation activity that happened at least
$2\times 10^8$~yrs before the time of observation and that the associated
object is in a quiescent phase. The UV radiation field is of the order of
or smaller than that in the Galaxy and the gas is cold ($T\sim 90-180$~K).
The $[$C\,{\sc i}$^\star /$C\,{\sc i}$]$ ratio in one of the components of the
system is consistent with what is expected for excitation by the CMBR field
at this redshift ($T=14.2$~K). This study demonstrates the importance of
this kind of observations for our understanding of how star formation proceeds
at high redshift. The prospect of using $\gamma$-ray burst afterglows as
background sources for similar studies is exciting and will boost this field
in a near future (Draine \& Hao 2002).

\acknowledgements{
This work is based on data collected under prog. 072.A-0346 of the
European Southern Observatory with UVES mounted on the 8.2~m Kueyen
telescope at the Paranal Observatory, Chile. We thank the referee,
Paolo Molaro, for fruitful comments. PP and RS acknowledge support from the
Indo-French Centre for the Promotion of Advanced Research (Centre Franco-Indien
pour la Promotion de la Recherche Avanc\'ee) under contract No. 3004-3.}

\clearpage
\begin{table}
\caption {Total column densities of different species in the DLA~system
at $z_{\rm abs}=4.224$ toward PSS J\,1443$+$2724}
\begin{tabular}{llcc}
\hline
\hline
Species & Transition & $\log N\pm\sigma _{\log N}$ & Metallicity                 \\
        & lines      &                             & relative to Solar $^{\rm a}$\\
\hline
H\,{\sc i}               & 1025,1215   & $20.95\pm 0.10$   & ...             \\
H$_2$                    & J=0 lines   & $17.67\pm 0.06$   & ...             \\
H$_2$                    & J=1 lines   & $17.97\pm 0.05$   & ...             \\
H$_2$                    & J=2 lines   & $17.48\pm 0.12$   & ...             \\
H$_2$                    & J=3 lines   & $17.26\pm 0.17$   & ...             \\
H$_2$                    & J=4 lines   & $14.02\pm 0.14$   & ...             \\
C\,{\sc i}               & 1277,1280   & $13.12\pm 0.02$   & ...             \\
C\,{\sc i}$^\star$       & 1277,1279   & $12.97\pm 0.03$   & ...             \\
C\,{\sc i}$^{\star\star}$& 1277,1280   & $\le 12.5$        & ...             \\
N\,{\sc i}               & 953.4,1134.1& $15.52\pm 0.01$   & $-$1.38$\pm$0.10\\
Mg\,{\sc ii}             & 1239,1240   & $15.98\pm 0.01$   & $-$0.55$\pm$0.10\\
S\,{\sc ii}              & 1250        & $15.52\pm 0.01$   & $-$0.63$\pm$0.10\\
Ar\,{\sc i}              & 1066        & $14.42\pm 0.01$   & $-$0.93$\pm$0.10\\
Fe\,{\sc ii}             & 1081        & $15.33\pm 0.03$   & $-$1.12$\pm$0.10\\
\hline
\end{tabular}
\label{Tabmetal}
\flushleft
$^{\rm a}$ Solar abundances adopted from Grevesse \& Sauval (2002).\\
Note: errors in the column densities of metals correspond to the rms
errors from fitting the Voigt profiles. They do not include the uncertainties
related to the continuum placement.
\end{table}

\clearpage
\begin{table}
\caption {Column densities and excitation temperatures from Voigt-profile
fitting of H$_2$ molecular lines}
\begin{tabular}{llcccc}
\hline
\hline
$z_{\rm abs}$ & Level & $\log N$              & $b\pm\sigma _b$ & $T_{\rm ex}$ & ${\rm J}_{\rm ex}$\\
              &       & $\pm\sigma _{\log N}$ & [km s$^{-1}$]   & [K]          &                   \\
\hline
4.22371 & J=0 & $17.41^{+0.02}_{-0.05}$ & $0.8\pm 0.3$     &    ...   & ...\\
        & J=1 & $17.59^{+0.03}_{-0.03}$ & \phantom{uu..}'' &   90-107 & 0-1\\
        & J=2 & $17.15^{+0.05}_{-0.13}$ & \phantom{uu..}'' &  200-259 & 0-2\\
        & J=3 & $16.53^{+0.17}_{-0.91}$ & \phantom{uu..}'' &  156-302 & 1-3\\
        & J=4 & $<13.6$ $^{\rm a}$      & \phantom{uu..}'' &    ...   & ...\\
4.22401 & J=0 & $17.34^{+0.09}_{-0.07}$ & $1.1\pm 0.3$     &    ...   & ...\\
        & J=1 & $17.75^{+0.06}_{-0.08}$ & \phantom{uu..}'' &  104-179 & 0-1\\
        & J=2 & $17.25^{+0.13}_{-0.14}$ & \phantom{uu..}'' &  218-377 & 0-2\\
        & J=3 & $17.21^{+0.13}_{-0.13}$ & \phantom{uu..}'' &  337-530 & 1-3\\
        & J=4 & $14.02^{+0.15}_{-0.07}$ & \phantom{uu..}'' &    ...   & ...\\
4.22416 & J=0 & $14.34^{+0.00}_{-0.12}$ & $1.7\pm 0.5$     &    ...   & ...\\
        & J=1 & $14.86^{+0.04}_{-0.13}$ & \phantom{uu..}'' &    ...   & ...\\
        & J=2 & $14.61^{+0.06}_{-0.15}$ & \phantom{uu..}'' &    ...   & ...\\
        & J=3 & $14.18^{+0.00}_{-0.16}$ & \phantom{uu..}'' &    ...   & ...\\
        & J=4 & $<13.6$ $^{\rm a}$      & \phantom{uu..}'' &    ...   & ...\\
\hline
\end{tabular}
\label{TabH2}
\flushleft
$^{\rm a}$ $3\sigma$ detection limit.\\
Note: errors in the column densities correspond to the range
$b\pm\sigma_b$ of Doppler parameters and not to the rms errors from fitting the
Voigt profiles.
\end{table}

\clearpage

\clearpage

\clearpage


\begin{thebibliography}{}

\bibitem[Ballester et al.(2000)]{2000Msg...101...31B} Ballester, P., Modigliani, A., Boitquin, O., Cristiani, S., Hanuschik, R., Kaufer, A., \& Wolf, S. 2000, The Messenger, 101, 31
\bibitem[Centuri\'on et al.(2003)]{2003A&A...403...55C} Centuri\'on, M., Molaro, P., Vladilo, G., P\'eroux, C., Levshakov, S.~A., \& D'Odorico, V. 2003, \aap, 403, 55
\bibitem[Cui et al.(2005)]{2005ApJ...633..649C} Cui, J., Bechtold, J., Ge, J., \& Meyer, D.~M. 2005, \apj, 633, 649
\bibitem[Dekker et al.(2000)]{2000SPIE.4008..534D} Dekker, H., D'Odorico, S., Kaufer, A., Delabre, B., \& Kotzlowski, H. 2000, \procspie, 4008, 534
\bibitem[D'Odorico \& Molaro(2004)]{2004A&A...415..879D} D'Odorico, V., \& Molaro, P. 2004, \aap, 415, 879
\bibitem[Draine \& Bertoldi(1996)]{1996ApJ...468..269D} Draine, B.~T., \& Bertoldi, F. 1996, \apj, 468, 269
\bibitem[Draine \& Hao(2002)]{2002ApJ...569..780D} Draine, B.~T., \& Hao, L. 2002, \apj, 569, 780
\bibitem[Ge \& Bechtold(1999)]{1999ASPC..156..121G} Ge, J., \& Bechtold, J. 1999, in Highly Redshifted Radio Lines, Carilli, C.~L., et al., eds., ASP Conf.~Ser. 156, 121
\bibitem[Grevesse \& Sauval(2002)]{2002ASR....30....3G} Grevesse, N., \& Sauval, A.~J. 2002, Adv. Space Res., 30, 3
\bibitem[Henry et al.(2000)]{2000ApJ...541..660H} Henry, R.~B.~C., Edmunds, M.~G., \& K\"oppen, J. 2000, \apj, 541, 660
\bibitem[Hirashita \& Ferrara(2005)]{2005MNRAS.356.1529H} Hirashita, H., \& Ferrara, A. 2005, \mnras, 356, 1529
\bibitem[Howk et al.(2005)]{2005ApJ...622L..81H} Howk, J.~C., Wolfe, A.~M., \& Prochaska, J.~X. 2005, \apjl, 622, L81
\bibitem[Izotov \& Thuan(1999)]{1999ApJ...511..639I} Izotov, Y.~I., \& Thuan, T.~X. 1999, \apj, 511, 639
\bibitem[Jura(1975)]{1975ApJ...197..581J} Jura, M. 1975, \apj, 197, 581
\bibitem[Ledoux(1999)]{1999PhDT........14L} Ledoux, C. 1999, Ph.D.~Thesis
\bibitem[Ledoux et al.(2002)]{2002A&A...392..781L} Ledoux, C., Srianand, R., \& Petitjean, P. 2002, \aap, 392, 781
\bibitem[Ledoux et al.(2003)]{2003MNRAS.346..209L} Ledoux, C., Petitjean, P., \& Srianand, R. 2003, \mnras, 346, 209
\bibitem[Levshakov \& Varshalovich(1985)]{1985MNRAS.212..517L} Levshakov, S.~A., \& Varshalovich, D.~A. 1985, MNRAS, 212, 517
\bibitem[Levshakov et al.(2002)]{2002ApJ...565..696L} Levshakov, S.~A., Dessauges-Zavadsky, M., D'Odorico, S., \& Molaro, P. 2002, \apj, 565, 696
\bibitem[Omont et al.(1996)]{1996Natur.382..428O} Omont, A., Petitjean, P., Guilloteau, S., McMahon, R.~G., Solomon, P.~M., \& P\'econtal, E. 1996, \nat, 382, 428
\bibitem[Petitjean et al.(2000)]{2000A&A...364L..26P} Petitjean, P., Srianand, R., \& Ledoux, C. 2000, \aap, 364, L26
\bibitem[Pettini et al.(2002)]{2002A&A...391...21P} Pettini, M., Ellison, S.~L., Bergeron, J., \& Petitjean, P. 2002, \aap, 391, 21
\bibitem[Prochaska et al.(2001)]{2001ApJS..137...21P} Prochaska, J.~X., et al. 2001, \apjs, 137, 21
\bibitem[Prochaska et al.(2002)]{2002AJ....123.2206P} Prochaska, J.~X., Gawiser, E., Wolfe, A.~M., Quirrenbach, A., Lanzetta, K.~M., Chen, H.-W., Cooke, J., \& Yahata, N. 2002, \aj, 123, 2206
\bibitem[Prochaska et al.(2003)]{2003ApJ...595L...9P} Prochaska, J.~X., Gawiser, E., Wolfe, A.~M., Castro, S., \& Djorgovski, S.~G. 2003, \apjl, 595, L9
\bibitem[Silva \& Viegas(2002)]{2002MNRAS.329..135S} Silva, A.~I., \& Viegas, S.~M. 2002, \mnras, 329, 135
\bibitem[Solomon \& Vanden Bout(2005)]{2005ARA&A..43..677S} Solomon, P.~M., \& Vanden Bout, P.~A. 2005, \araa, 43, 677
\bibitem[Vladilo et al.(2003)]{2003A&A...402..487V} Vladilo, G., Centuri\'on, M., D'Odorico, V., \& P\'eroux, C. 2003, \aap, 402, 487
\bibitem[van Zee et al.(1998)]{1998AJ....116.2805V} van Zee, L., Salzer, J.~J., Haynes, M.~P., O'Donoghue, A.~A., \& Balonek, T.~J. 1998, \aj, 116, 2805

\end{thebibliography}
\end{document}